\documentstyle[12pt]{article}

\newcommand{\nummer}[1]{\hskip 12 true cm #1 \par}
\newcommand{\monat}[1]{\vspace{-14pt}\hskip 12 true cm #1
                       \par \vspace*{1 cm}}
\newcommand{\titel}[1]{{\renewcommand{\thefootnote}{\fnsymbol{footnote}}
                       \Large\bf\vskip 0 true cm
                       \begin{center}#1\end{center}
                       \setcounter{footnote}{0}}
                       \normalsize\vskip 1.2 true cm}
\newcommand{\autor}[1]{{\renewcommand{\thefootnote}{\fnsymbol{footnote}}
                       \begin{center} {\large #1 }\end{center}}
                       \setcounter{footnote}{0}}
\newcommand{\adresse}[1]{\vspace*{-1.1 true cm}\begin{center} {\it #1 }
                         \end{center}
                         \vskip 0.5cm}

\newcommand{\pslash}{\kern 0.1 em p\kern -0.45em /}
\newcommand{\dslash}{\kern 0.1 em \partial\kern -0.55em /}
\newcommand{\sla}[1]{\kern 0.1 em #1\kern -0.55em /}

\newcommand{\lra}{\longrightarrow}

\newcommand{\RA}{\Rightarrow}

\newcommand{\R}{\mbox{I\kern -0.22em R\kern 0.30em}}
\newcommand{\N}{\mbox{I\kern -0.22em N\kern 0.30em}}
\newcommand{\C}{\mbox{\kern 0.20em \raisebox{0.09ex}{\rule{0.08ex}{1.22ex}}
                \kern -0.60em C\kern 0.30em}}
\newcommand{\Z}{\mbox{\sf Z\kern -0.40em Z\kern 0.30em}}

\newtheorem{theo}{Theorem}
\newtheorem{prop}{Proposition}
\newtheorem{lemm}{Lemma}
\newtheorem{defi}{Definition}

\def\sss{\scriptscriptstyle}
\def\OD{\Omega_{\scriptscriptstyle D}}
\def\cA{\cal A}
\def\cAM{\cA_{\scriptscriptstyle M}}
\def\cF{\cal F}
\def\cH{\cal H}
\def\cJ{\cal J}
\def\cM{\cal M}
\def\cL{\cal L}
\def\dD{d_{\scriptscriptstyle D}}
\def\dM {d_{\scriptscriptstyle M}}
\newcommand{\cl}[1]{\mbox{[}#1\mbox{]}}

\newcommand{\qquotient}[2]{{ #1\over #2}}

\newcommand{\xkern}{\mbox{ker}}

\catcode`\@11
\def\section{
    \setcounter{equation}{0}
\@startsection {section}{1}{\z@}{-3.5ex plus -1ex minus -.2ex}
{2.3ex plus .2ex}{\large\bf}
}
\catcode`\@13

\begin{document}


\nummer{MZ-TH/93-28}
\monat{October 1993}
\titel{Differential Algebras in Non-Commutative Geometry}
\autor{W.~Kalau, N.A.~Papadopoulos, J.~Plass, J.-M.~Warzecha}
\adresse{Johannes Gutenberg Universit\"at\\
Institut f\"ur Physik\\
55099 Mainz}
\begin{abstract}
We discuss the differential algebras used in Connes' approach to
Yang-Mills theories with spontaneous symmetry breaking. These differential
algebras generated by algebras of the form
functions $\otimes$ matrix are shown to be skew tensor products of
differential forms with a specific matrix algebra. For that we derive a general
formula for differential algebras based
on tensor products of algebras. The result is used to characterize
differential algebras which appear in models with one symmetry breaking
scale.
\end{abstract}
\newpage


\section{Introduction}

In spite of the great experimental success of the Standard Model of
electroweak interaction there is a general feeling that the theoretical
understanding of this interaction is far from being complete. Not only
that the regularities, like the appearance of the elementary particles in
families, and the irregularities, like the mass-matrix, are a complete mystery.
Also the concept of spontaneous symmetry breaking seems to be arbitrarily
introduced by hand in order to turn Yang-Mills theories into experimentally
relevant models.

However, there are recent new and promising attempts to solve at least the
problem related to the Higgs-mechanism and spontaneous symmetry breaking.
They are more or less based on or inspired by Connes' non-commutative geometry
\cite{cobuch}. There is one line of approach initiated by Connes himself
\cite{CoLo} which lateron was generalized to a Grand Unification Model
\cite{chams,chams1}.

There is another line of approach followed in \cite{robert,florian}. It is
based on superconnections in the sense of Matthai, Quillen \cite{MaQui}.
The key-idea in these models is to extend the usual exterior differential
by a matrix derivation. The physical motivation for this was given in
\cite{hps2}. The connection is then taken to be an element of
a graded Lie-algebra $SU(m|n)$ which has been extended to a module over
differential forms. This line of approach seems to be related to the one
of Connes. However, until now it was not known to what extent they are
similar and
what the precise differences are. A precise comparison between the models
\cite{CoLo,kastler} and \cite{robert,florian,hps2} became possible only
after the
present construction. The results of this comparison will appear in another
publication \cite{pps}.

In this article we want to investigate Connes' approach to Yang-Mills
theory with spontaneous symmetry breaking. More precisely, we will discuss
the differential algebra $\OD\cA$ in Connes construction.
This algebra is a derived
object, obtained from an associative algebra $\cA$ via the universal
differential enveloping algebra and a k-cycle. Therefore $\OD\cA$ is not
known in general and has to be computed for each specific example,
we shall give a quite detailed characterization of this object for general
situations.
The fact that all physical quantities like connections or curvatures are
objects in
the differential algebra $\OD\cA$ underlines its importance. Some attempts
for the construction of the algebra $\OD\cA$ were done by \cite{KT}.
A.Connes gives some special examples in \cite{cobuch}.

We show that this algebra is in fact a skew tensor product of a specific
differential
matrix algebra with differential forms, i.e. matrix valued differential forms.

In the case of Yang-Mills theories with spontaneous symmetry breaking the
algebra $\cA$ is given as a tensor product of $\cF$, the algebra of smooth
functions, and $\cAM$, a matrix algebra. The differential algebra $\OD\cF$
for the algebra of functions is the usual de Rham algebra \cite{cobuch}.
$\OD\cAM$ for the matrix algebra will be easy to compute, as we shall see
in sec.~3. Therefore we want to make use of this fact and derive in sec.~4 a
general formula which relates $\OD(\cA_1\otimes\cA_2)$ of a product algebra
to the differential algebras $\OD\cA_1$ and $\OD\cA_2$ of the factor algebras.
In our case, where the tensor product of an algebra of functions and a
matrix algebra
is taken, the general relation becomes much simpler. Thus it is straightforward
 to write down the complete algebra $\OD(\cF\otimes\cAM)$ which we
will do for the 2-point case in sec.~5. This article ends with conclusions
drawn
in sec.~6. However, we shall first give a brief introduction to the general
subject in the next section.


\section{The Universal Differential Envelope and k-Cycles}

We start with a brief review of the basic concepts of non-commutative
geometry needed to describe Yang-Mills theories with spontaneous symmetry
breaking. This will allow us to fix the notation and to introduce some
useful definitions. For a more comprehensive presentation of this subject
we refer to \cite{cobuch,josch,jan}.

Let $\cA$ be an associative unital algebra. We can construct a bigger algebra
$\Omega\cA$ by associating to each element $A\in\cA$ a symbol
$\delta A$. $\Omega\cA$ is the free algebra generated by the symbols $A$,
$\delta A$, $A\in\cA$ modulo the relation
\begin{equation}
   \delta (AB)=\delta A\, B + A\delta B\;\; .\label{gl-4}
\end{equation}
With the definition
\[
   \delta(A_0\delta A_1\cdots\delta A_k)
   \;:=\;
   \delta A_0\,\delta A_1\cdots\delta A_k
\]
$\Omega\cA$ becomes a $\N$-graded differential algebra with the odd
differential $\delta$, $\delta^2=0$. $\Omega\cA$ is called the universal
differential envelope of \cA.

The next element in this formalism  is a k-cycle $(\cH, D)$ over $\cA$, where
$\cH$ is a Hilbert space such that there is an algebra homomorphism
\[
\pi : \cA \lra B(\cH)\;\; .
\]
$B(\cH)$ denotes the algebra of bounded operators acting on $\cH$. $D$ is a
Dirac operator such that $[D,\pi(A)]$ is bounded for all $A\in \cA$. We
can use this operator to extend $\pi$ to an algebra homomorphism of $\Omega\cA$
by defining
\[
\pi(A_0\delta A_1\cdots\delta A_k):=\pi(A_0)[D,\pi(A_1)]\cdots[D,\pi(A_k)]
\;\; .
\]
However, in general $\pi(\Omega\cA)$ fails to be a differential algebra.
In order to repair this, one has to divide out the two sided $\N$-graded
differential ideal $\cJ$ given by
\begin{equation} \label{gl-1}
   \cJ := \bigoplus \cJ^k\;\; ,\;\; \cJ^k:=K^k+\delta K^{k-1}\;\;,\;\;
         K^k:=\hbox{ker}\pi\cap\Omega\cA^k \;.
\end{equation}
Now we are ready to define our basic object of interest, $\OD$ as
\[
   \OD := \bigoplus_{k\in\N} \pi(\Omega\cA^k) /\pi(\cJ^k) \;\;.
\]
$\OD$ is an $\N$-graded differential algebra, where the differential $d$ is
defined by
\[
   d[\pi(\omega)] := [\pi(\delta\omega)]\;\;, \;\; \omega\in\Omega\cA\;\;.
\]
If we take, for example,
$\cA = \cF$, the algebra of smooth functions on a compact
spin-manifold, the space of square-integrable spin-sections as $\cH$ and
$D=i\dslash$ then $\OD$ is the usual de Rham-algebra \cite{cobuch}.

Since we are dealing with tensor products of algebras and want to make use
of this
fact we also have to extend k-cycles over factor algebras to a k-cycle of
product algebras. One possibility is provided by the notion of product
k-cycles.
Suppose we have
two k-cycles, $(D_1,\cH_1)$ over $\cA_1$ , $(D_2,\cH_2)$ over $\cA_2$ and
suppose
there is a $\Z_2$-grading on $\cH_1$ given by $\Gamma_1$. The product of
k-cycles $(D_{12},\cH_{12})$ over $\cA=\cA_1\otimes\cA_2$ is given as
\begin{eqnarray}
   \cH_{12} &:=& \cH_1 \otimes \cH_2 \nonumber \\
   D_{12} &:=& D_1\otimes 1 + \Gamma_1\otimes D_2 \;\; . \label{gl-26}
\end{eqnarray}


\section{Differential Forms of Associative Matrix Algebras}

In this section we shall derive some general properties of the differential
algebra generated by an associative matrix algebra $\cA$ . The k-cycle
($\cH$ , D) over $\cA$ is specified as
\begin{equation}
\cH = \C^N \;\; ;\;\; D:= [\cM, \cdot]\; ,\;\; \cM\in \C^{\sss N\otimes N}\; .
\label{gl-27}
\end{equation}
Without any loss of generality we may assume that for the algebra
homomorphism
\[
   \pi : \cA \lra \C^{\sss N\otimes N}
\]
we have
\[
   \hbox{ker}\left(\pi(\cA)\right) = \{ 0\} \; .
\]
Obviously in this case $\cJ^1$ is given by
\[
   \cJ^1= \xkern\left(\pi(\Omega\!\cA)\right) \cap \Omega^1\!\cA\; .
\]
Therefore we conclude that the first non-trivial contributions of the
differential operator can only appear at degree $k\geq 2$ of $\cJ$.

For quite general cases the
next lemma shows that $\cJ$ is generated by $\cJ^2$ and as a consequence
the differential on $\OD$ is given by a supercommutator.
\begin{lemm}
Let $\cA$ be an associative algebra and ($\cH$, D) a K-cycle over $\cA$ as in
(\ref{gl-27}) and $\pi$ the corresponding algebra homomorphism with
$\hbox{ker}\left(\pi(\cA)\right) = \{ 0\}$ .
If
\begin{equation}
   [\cM^2,\pi(\cA)] \subset \pi(\cJ^2) \label{gl-25}
\end{equation}
then
\begin{itemize}
\item[i)]
$\pi(\cJ)$ is generated by $\pi(\cJ^2)$, i.e.
\begin{equation}
   \pi(\cJ^k) \;=\; \sum_{j=0}^{k-2}
     \pi\big(
       \Omega^j\!\cA \, \cJ^2 \, \Omega^{k-j-2}\!\cA
     \big)
   \;\; , \;\; k\geq 2\;\; ; \label{gl-24}
\end{equation}
\item[ii)]
the differential $d$ on $\OD\!\cA$ is given by the graded supercommutator
\[
   d[\omega^k] = \left[ [\cM,\pi (\omega^k)]_{\sss S} \right] \;=\;
   \left[ \cM\pi (\omega^k) - (-1)^k\pi (\omega^k)\cM\right]\;\;
\]
with $[\omega^k]\in\OD^k\!\cA$ and $\omega^k\in\Omega^k\!\cA\;$.
\end{itemize}
\end{lemm}

{\bf Proof:} Let us consider
$\pi(\omega^k)= [\cM ,a_1] \cdots [\cM, a_k]$ with
$a_1[\cM ,a_2] \cdots [\cM , a_k] = 0\;$, $ a_i\in\pi(\cA)$ ,
i.e. $\omega^k \in \cJ^k$. We have
\begin{eqnarray*}
   \pi(\omega^k) &=& [\cM ,a_1] \cdots [\cM, a_k]
   \;=\; -a_1\cM[\cM ,a_2]\cdots [\cM ,a_k]  \\
   &=& -a_1[\cM^2 ,a_2] \cdots [\cM, a_k] + a_1[\cM ,a_2]\cM\cdots [\cM ,a_k]\\
   &\vdots& \\
   &=& -\sum_{j=2}^{k} (-1)^j a_1[\cM , a_2]\cdots [\cM^2 , a_j]\cdots
       [\cM ,a_k]
       + (-1)^k a_1[\cM ,a_2] \cdots [\cM , a_k]\cM  \;\;.
\end{eqnarray*}
The last term in the last equation vanishes by assumption and the sum is of
the form as in eq.(\ref{gl-24}) which proves {\it i)}.

The second part of the lemma is proved by a similar calculation.
We now take $\pi(\omega^k)= a_0[\cM ,a_1] \cdots [\cM, a_k]$ as a
representative
of $\cl{\omega^k}\in \OD\!\cA$. We have to show that the differential
\begin{equation}
   d\cl{\omega^k} = \cl{ \cM\pi (\omega^k) - (-1)^k\pi (\omega^k)\cM}
   \label{gl-23}
\end{equation}
coincides with the one induced by the differential on the universal
differential envelope of $\cA$.
The representative of the first term of the right hand side of eq.(\ref{gl-23})
can be rewritten as
\begin{eqnarray*}
  \cM\pi(\omega^k)
   &=&  \cM a_0 [\cM ,a_1] \cdots [\cM, a_k] \\
 &=& [\cM ,a_0][\cM ,a_1]\cdots [\cM ,a_k] + a_0 \cM[\cM ,a_1] \cdots
[\cM , a_k] \\
  &\vdots& \\
  &=& [\cM ,a_0][\cM ,a_1]\cdots [\cM ,a_k]
  -\sum_{j=1}^{k} (-1)^j a_1[\cM , a_2]\cdots [\cM^2 , a_j]\cdots [\cM ,a_k]
\\
  && +\; (-1)^k a_1[\cM ,a_2] \cdots [\cM , a_k]\cM  \;\; .
\end{eqnarray*}
Thus we see that
\[
\cM\pi (\omega^k)] \;=\; [\pi (\delta\omega^k)] + (-1)^k[\pi(\omega^k)\cM]\;\;
{}.
\]
Inserting this in eq.(\ref{gl-23}) we obtain
\[
   d\cl{\omega^k} \;=\; \cl{ \pi (\delta\omega^k)}\; .
\]
Since $\cJ$ is a differential ideal the result is independent of the choice
of a representative and we have proved the lemma.

Our next task is to find algebras for which the condition (\ref{gl-25})
is fulfilled. The next lemma shows that the matrix algebras
which are building blocks in models discussed in \cite{CoLo} for the two
point case and in \cite{chams} for the n-point case meet condition
(\ref{gl-25}). They all have in common that they are  direct sums of
algebras $\cA=\cA_1 \oplus \cdots \oplus\cA_n$
such that the algebra homomorphism maps them into a block diagonal
matrix of the form
\[
\pi(\cA) = \left(
\begin{array}{ccc}
\pi_1(\cA_1)& & \\
 &\ddots& \\
 & &\pi_n(\cA_n)
\end{array} \right)
\]
where $\pi_j$ denotes the restriction of $\pi$ to $\cA_j$. The Dirac operator
for those algebras is off-diagonal which will be made more precise in the
following lemma.

\begin{lemm}
Let $\cA$ be an associative algebra which can be decomposed into a direct
sum \linebreak
$\cA = \cA_1 \oplus \cdots \oplus\cA_n$ of unital algebras and ($\cH$,D)
a K-cycle over $\cA$ as in (\ref{gl-27}). $P_i$,
\newline $i=1,\dots , n$ are
projection operators on $\cH$ with
\[
   P_i\pi(\cA_i)= \pi(\cA_i)=\pi(\cA_i)P_i  \;\; ; \;\;
   P_iP_j = \delta_{ij}                     \;\; ; \;\;
   \sum_i P_i = 1 \;\;.
\]
If
\begin{equation}
   P_i\cM P_i = 0 \;, \; i= 1, \dots , n \label{gl-22}
\end{equation}
then
\[
   [\cM^2 , \pi(\cA)] \subset \pi(\cJ^2)\; .
\]
\end{lemm}

{\bf Proof:} We introduce the following notation
\begin{equation}
   \cM_{ij} := P_i\cM P_j \;\; ,\;\;
   a^i:= P_i\pi(A)\;\; , \;\;
   b^i:= P_i\pi(B)\;\; ,\; A,B\in\cA\;\; . \label{gl-21}
\end{equation}
Elements $\pi(\omega^2) \; ,\; \omega^2 \in \cJ^2$ can be written as
\begin{equation}
   \pi(\omega^2) =  \pi(A)[\cM^2 , \pi(B)] \label{gl-19}
\end{equation}
where $A , B$ obey the condition
\begin{equation}
   \pi(A)[\cM , \pi(B)] = 0\;\; . \label{gl-20}
\end{equation}
In the notation introduced in (\ref{gl-21}) eqs. (\ref{gl-19},\ref{gl-20})
take the form
\begin{eqnarray*}
   P_i\pi(\omega^2)P_j&=& \sum_{k=1}^{n} a^i\cM_{ik}\cM_{kj}b^j -
   \sum_{k=1}^{n}a^ib^i\cM_{ik}\cM_{kj}\\
   0 &=& a^i\cM_{ij}b^j - a^ib^i\cM_{ij}\;\; .
\end{eqnarray*}
For any $A^\prime_i \in \cA_i$ one can choose
\begin{equation}
   \begin{array}{rcl}
      A_1 = A^\prime_i \;& , &  \; B_1= {\bf 1_i}\\
      A_2 = -{\bf 1_i}  \;& , &\;  B_2 = A^\prime_i\;\; .
   \end{array} \label{gl-18}
\end{equation}
Since $\cM_{ii} = 0$ it is straightforward to verify that
\[
   \sum_{r\in\{1,2\} } \pi(A_r)[\cM , \pi(B_r)] = 0
\]
and
\[
   \sum_{r\in\{1,2\} } \pi(A_r)[\cM^2 , \pi(B_r)] =
   [\cM^2, \pi(A^\prime_i)]\;\;  .
\]
This shows that
\[
   [\cM^2, \pi(\cA_i)]\subset \pi(\cJ^2)\;\;  .
\]
Since elements $A_r,B_r\in\cA$ as in eq.(\ref{gl-18}) exist for any
$i\in\{1,\dots ,n\}$ the lemma is proved.

We now want to apply these results to a two point case, i.e. to a matrix
algebra which is given as the direct sum of the algebras
$\cA_1=\C_{\sss n\times n}$
and $\cA_2=\C_{\sss m\times m}$ of complex
$n\times n$ resp.~$m\times m$ matrices. The representation of the algebra
and the Dirac operator take the form
\[
\pi(\cA) = \left(
\begin{array}{cc}
\cA_1& 0 \\
0 &\cA_2
\end{array} \right)\;\;\; ,\;\;\;
\cM = \left(
\begin{array}{cc}
0 & \mu^* \\
\mu & 0
\end{array} \right)
\]
where $\mu$ denotes an arbitrary (non-zero) complex $n\times m$ matrix.
Let us first consider the algebra generated by
\begin{equation}
   \cA_1 [\mu^*\mu ,\cA_1]\cA_1\;\; . \label{gl-17}
\end{equation}
There are two possibilities. Either $\mu^*\mu \sim 1_{m\times m}$, then the
commutator in (\ref{gl-17}) is zero and no non-trivial algebra can be
generated, or $\mu^*\mu \sim\!\!\!\!\!\!/\; 1_{m\times m}$, then the whole
algebra $\cA_1$ is generated. There is the same situation for
\[
\cA_2 [\mu\mu^*,\cA_2]\cA_2
\]
and therefore we may distinguish three cases:
\begin{itemize}
\item[{\bf i.}] $\mu^*\mu \sim 1_{m\times m}$ and $\mu\mu^* \sim 1_{n\times n}$
which
is possible only for $m=n$, i.e $\cA_1=\cA_2$. In this case we have
$\cJ= \{0\}$ and
\begin{equation}
   \OD^{2n}\cA =
   \left( \begin{array}{cc}
          \cA_1& 0\\
          0&\cA_1
   \end{array}\right)
   \;\;\; ,\;\;\;
   \OD^{2n+1}\cA =
   \left( \begin{array}{cc}
          0 & \cA_1\\
          \cA_1& 0
   \end{array}\right) \;\; , n\in\N\;\; .\label{gl-16}
\end{equation}
The multiplication is just the ordinary matrix multiplication of
$2m\times 2m$ matrices.
\item[{\bf ii.}] $\mu^*\mu \sim\!\!\!\!\!\!/\; 1_{m\times m}$ and
$\mu\mu^* \sim\!\!\!\!\!\!/\; 1_{n\times n}$. Here only $\OD^1\cA$ survives
since
\[
\pi(\Omega^2\cA) = \pi(\cA) = \cJ^2
\]
and therefore
\[
\pi(\Omega^k\cA) = \cJ^k\;\; k\geq 2\;\;.
\]
In this case one may view
\[
   \OD^1\cA =
   \{ A\in \C^{m\times n} \} \oplus
   \{ B\in \C^{n\times m} \}
\]
as a module over $\cA$. There is no non-trivial multiplication of elements
in $\OD^1\cA$, i.e. for $\nu,\omega\in\OD^1\cA$ we have $\nu\cdot\omega=0$.
\item[{\bf iii.}] $m \leq n$, $\mu^*\mu \sim 1_{m\times m}$ and
$\mu\mu^* \sim\!\!\!\!\!\!/\; 1_{n\times n}$. Again we have
\[
   \OD^1\cA =
   \{ A\in \C^{m\times n} \} \oplus
   \{ B\in \C^{n\times m} \}
\]
as a module over $\cA$. However, in this case $\OD^2\cA$ is non-trivial
since
\[
J^2=\left(\begin{array}{cc}
0& 0\\
0 & \cA_2\end{array}\right)\;\; \RA\;\;
\pi(\Omega^k\cA) = \cJ^k\;\; k\geq 3\;\;.
\]
and therefore
\[
\OD^2\cA = \left( \begin{array}{cc}
\cA_1& 0\\
0 & 0\end{array}\right)
\]
and all higher degrees of $\OD\cA$ are trivial. The multplication $\circ$ of
two elements $(A,B), (A^\prime,B^\prime) \in \OD^1\cA$ is given by
\[
(A,B)\circ (A^\prime,B^\prime) =
\left(\begin{array}{cc}
A\cdot B^\prime& 0 \\
0 & 0
\end{array}\right)
 \in \OD^2\cA
\]
where $\cdot$ denotes the usual matrix multiplication.
A representation for the matrix algebra is now on the zeroth and first
degree given by
\[
\OD^0\cA = \left(\begin{array}{cc}
\cA_1 & 0 \\
0   & \cA_2
\end{array}\right)
\;\;\; ,\;\;\;
\OD^1\cA = \left(\begin{array}{cc}
 0                    & \eta \C^{m\times n} \\
\eta^\prime\C^{n\times m}   & 0
\end{array}\right)\;\; .
\]
The relations for the formal elements $\eta$, $\eta^\prime$ are
\[
   \eta\eta^\prime\neq 0 \;\; ,\;\; \eta^\prime\eta =0
   \;\; .
\]
Although these relations seem a little  awkward it is not difficult to
find a representation for them. E.g.
\[
\eta=\left(\begin{array}{ccc}
 0 & 1 \\
 0 & 0
\end{array}\right) \;\;\; ,\;\;\;
\eta^\prime=\left(\begin{array}{ccc}
 0 & 0 \\
 0 & 1
\end{array}\right)
\]
is such an representation.

\end{itemize}


\section{$\OD$ of the Tensor Product of Algebras}

In this section we are going to establish a relation allowing the computation
of
$\OD$ for a tensor product of two
algebras $\cA = \cA_1\otimes \cA_2$. The idea is to use the knowledge of
$\OD(\cA_i)$ in order to calculate $\OD(\cA)$.
This cannot be done by simply taking tensor products of $\OD\cA_1$ and
$\OD\cA_2$. We observe that already at the level of universal differential
envelopes we have in general
\[
   \Omega(\cA_1\otimes\cA_2) \neq \Omega\cA_1\hat\otimes\Omega\cA_2\;
   \footnote{The hat on the tensor product denotes the $\Z_2$ graded tensor
   product.}\; .
\]
Let us denote by $\delta_{12}$ the differential on $\Omega(\cA_1\otimes\cA_2)$.
The differential on $\Omega\cA_1\hat\otimes\Omega\cA_2$ is
$\delta_S:= \delta_1\otimes 1 +\chi\otimes\delta_2$, where $\delta_i$ denotes
the differentials on $\Omega\cA_i$ and $\chi$ is the $\Z_2$ grading on
$\Omega\cA_1$. We apply $\delta_S$ on some arbitrary
element $(a\otimes b)\in\cA_1\otimes\cA_2$
\begin{eqnarray*}
  \delta_S(a\otimes b)
  &=&   \delta_1 a\otimes b + a\otimes \delta_2 b \\
  &=& (1\otimes b)(\delta_1 a\otimes 1)+(a\otimes 1)(1\otimes\delta_2 b)\\
  &=&   (1\otimes b)\delta_S (a\otimes 1)+(a\otimes 1)\delta_S(1\otimes b)\;\;
{}.
\end{eqnarray*}
For $\delta_{12}$ we can only use relation (\ref{gl-4}) to write
\[
   \delta_{12}(a\otimes b) = \delta_{12}(a\otimes 1)\, (1\otimes b)
   +(a\otimes 1)\delta_{12}(1\otimes b)\;\; ,
\]
there is no relation which tells us that
\[
   \delta_{12}(a\otimes 1)\, (1\otimes b) - (1\otimes b)\delta_{12}(a\otimes 1)
   =0\;\; .
\]
Thus we see that
$\Omega(\cA_1\otimes\cA_2)$ and $\Omega\cA_1\hat\otimes\Omega\cA_2$ are not
isomorphic.

However, we can prove the following lemma.
\begin{lemm} \label{lemm-2}
   Let $(\omega_1,\delta_1)$ and $(\omega_2,\delta_2)$ be two graded
  differential algebras which are both
  generated by the same algebra ${\cA}$ as zeroth grading and their
  respective differentials. Let ${\cal B}$ be an algebra and let
   \begin{eqnarray*}
        \pi_1 : \omega_1&\rightarrow& {\cal B}\\
        \pi_2 : \omega_2&\rightarrow& {\cal B}
   \end{eqnarray*}
   be algebra homomorphisms. $\cJ_{\pi_i}$ are the
   corresponding differential ideals as defined in (\ref{gl-1}) and
$\Omega_{\pi_i} = \bigoplus_{\sss k\in N} \omega^k_i/J^k_{\pi_i}$ the induced
   differential algebras.

  If
   \[
   \pi_1(a)=\pi_2(a) \;\;\mbox{and}\;\; \pi_1(\delta_1a)=\pi_2(\delta_2a)
  \;\; \mbox{for all}\;\; a\in {\cA}
   \]
   then
   \[
    \Omega_{\pi_1}=\Omega_{\pi_2}\;\; .
   \]
   \end{lemm}
{\bf Proof:}
   As $\omega_1$ and $\omega_2$ are generated by elements
   $a_0\delta_ia_1 \ldots \delta_ia_k\; i=1,2$,  we obviously have
   \[
      \pi_1(\omega_1) = \pi_2(\omega_2)
   \]
   since $\pi_i$ are algebra homomorphisms.
Therefore the relation $a_0\delta_1a_1 \ldots \delta_1a_k \in \xkern^k_{\pi_1}$
   implies $a_0\delta_2a_1 \ldots \delta_2a_k \in \xkern^k_{\pi_2}$
   and we get
   \[
   \pi_1(J_{\pi_1})=\pi_2(J_{\pi_2}).
   \]
    This establishes the identity of subalgebras of $\cal B$.

An example for this situation is $\Omega({\cA_1\otimes \cA_2})$ and
$\Omega({\cA_1})\hat{\otimes} \Omega({\cA_2})$. The lemma allows to
use the second algebra for the calculation of $\OD(\cA_1\otimes\cA_2)$ because
of our choice of the product k-cycle. Thus we now have to analyze
\[
   \qquotient{\pi\big( (\Omega\cA_1\hat{\otimes} \Omega\cA_2)^k\big)}
             {\pi(\delta\xkern \pi^{k-1}+\xkern^k\pi)} \;\;.
\]
This can be done by splitting the mapping
\[
   \pi=\pi_1\otimes\pi_2 :\;\;\Omega{\cA_1}\hat{\otimes} \Omega{\cA_2}
   \;\longrightarrow\;
   {B(\cH_1\otimes \cH_2)}
\]
into $\pi=\pi_\Sigma\circ\pi_\oplus$ with $\pi_\oplus$ defined by:
\[
   \begin{array}{rccc}
      \pi_\oplus:& (\Omega\cA_1\hat{\otimes} \Omega\cA_2)^k&
\rightarrow &
\bigoplus_{i+j=k}\pi_1(\Omega{\cA_1}^i)\otimes\pi_2(\Omega{\cA_2}^j)\\[1ex]
      &a_0\delta_1a_1 \ldots \delta_1a_i\otimes b_0\delta_2b_1\ldots
\delta_2b_j
      &\mapsto&
      \pi_1(a_0\delta_1a_1 \ldots \delta_1a_i)\otimes
      \pi_2(b_0\delta_2b_1\ldots \delta_2b_j)
   \end{array}
\]
Accordingly $\pi_\Sigma$ is then given by the summation:
\[
   \begin{array}{rccc}
      \pi_\Sigma:&\bigoplus_{i+j=k}\pi_1(\Omega{\cA_1}^i)\otimes
      \pi_2(\Omega{\cA_2}^j)&\rightarrow & {B(\cH_1\otimes \cH_2)} \\
      & \Sigma_{i+j=k}  (\nu_i\otimes\omega_j) & \mapsto & \sum_{i+j=k}\nu_i
      \otimes\omega_j
   \end{array}
\]
I.e. $\pi_\oplus$ maps each term of the sum
\[
   (\Omega\cA_1\hat\otimes\Omega\cA_2)^k = \bigoplus_{i+j=k}
   \Omega\cA_1^i\otimes\Omega\cA_2^j
\]
to a separate copy of $B(\cH_1\otimes\cH_2)$, in a second step
these copies are identified by $\pi_\Sigma$.

The reason for this splitting will become clearer by calculating the
quotient
\[
   \qquotient{\pi_\oplus\big((\Omega\cA_1\hat{\otimes} \Omega\cA_2)^k\big)}
             {\pi_\oplus(\delta\xkern \pi_\oplus^{k-1}+\xkern^k\pi_\oplus)}
\]
We fix the notation:
\[
   J_\oplus^k = \delta\xkern \pi_\oplus^{k-1} + \xkern\pi_\oplus^k
\]
With the following result from multilinear algebra
\[
   \xkern\pi_\oplus^k=\bigoplus_{i+j=k} \xkern\pi_1^i\otimes\Omega{\cA_2}^j+
   \Omega{\cA_1}^i\otimes\xkern\pi_2^j\; ,
\]
this gives
\[
   \pi_\oplus(J_\oplus^k)=\bigoplus_{i+j=k}\pi_1(J_{\pi_1}^i)\otimes\pi_2
   (\Omega{\cA_2}^j) +\pi_1(\Omega{\cA_1}^i)\otimes\pi_2(J_{\pi_2}^i)
\]
with the abbreviation:
\[
   \pi_i(\delta_i\xkern\pi_i^{j-1})=\pi_i(J_{\pi_i}^j)
\]
There is now:
\begin{eqnarray}
   \qquotient{\pi_\oplus(\Omega(\cA_1\otimes\Omega\cA_2)^k)}
             {\pi_\oplus(J_\oplus^k)}
   &=&
   \qquotient{ \bigoplus_{i+j=k}\pi_1(\Omega{\cA_1}^i)
               \otimes\pi_2(\Omega{\cA_2}^j)
             }
             { \bigoplus_{i+j=k}\pi_1(J_{\pi_1}^i)\otimes\pi_2
               (\Omega{\cA_2}^j)+\pi_1(\Omega{\cA_1}^i)
               \otimes\pi_2(J_{\pi_2}^j)
	     } \nonumber \\
   &=&
   \bigoplus_{i+j=k}
   \qquotient{\pi_1(\Omega{\cA_1}^i)\otimes\pi_2(\Omega{\cA_2}^j)}
             {\pi_1(J_{\pi_1}^i)\otimes\pi_2(\Omega{\cA_2}^j)+
	        \pi_1(\Omega{\cA_1}^i)\otimes\pi_2(J_{\pi_2}^j)
	     }\nonumber\\
   &=&
   \bigoplus_{i+j=k}
   \qquotient{\pi_1(\Omega{\cA_1}^i)}{\pi_1(J_{\pi_1}^i)}
   \otimes
   \qquotient{\pi_2(\Omega{\cA_2}^i)}{\pi_2(J_{\pi_2}^i)}\nonumber\\
   &=&
   \bigoplus_{i+j=k} \OD\cA_1^i\otimes\OD\cA_2^j
   \label{gl-15}
\end{eqnarray}
The multiplication is here given by the skewsymmetric multiplication
\[
   \nu_1\otimes\omega_1\cdot\nu_2\otimes\omega_2=
   (-1)^{\partial\omega_1\partial\nu_2}
   \nu_1\nu_2\otimes\omega_1\omega_2
\]
and the differential is :
\[
   d(\nu\otimes\omega)=d_1\nu\otimes\omega+(-1)^{\partial\nu}
   \nu\otimes d_2\omega
\]
Thus we see that the differential algebra related to $\pi_\oplus$ is
just the tensor product
\begin{equation}
   \OD\cA_1\hat{\otimes}\OD\cA_2\;\; .\label{gl-14}
\end{equation}
Suppose the $\OD\cA_i$ are known. The non-trivial step in the calculation
of $\OD(\cA_1\otimes\cA_2)$ is then related to the map $\pi_\Sigma$.
Before we formulate the proposition relating the tensor product
(\ref{gl-14}) to $\OD(\cA_1\otimes\cA_2)$ via $\pi_\Sigma$
we need one further definition:
\begin{defi}\label{defi-1}
If $\beta: \omega \rightarrow \beta(\omega)$ is an algebra homomorphism
and $j$
   an ideal in $\omega$ then let $\beta_q$ be the homomorphism
   \[
     \beta_q:\;\; \qquotient{\omega}{j}\rightarrow \qquotient{\beta(\omega)}
     {\beta(j)} \;\;.
   \]
\end{defi}

\begin{prop}\label{prop-1}
	Let $(\omega,d),(\omega_1,d_1),(\omega_2,d_2)$ be  $N$-graded
        differential algebras, $\alpha$
	and $\beta$ differential algebra homomorphisms
        \[
          \omega   \;\stackrel{\alpha}{\longrightarrow}\;
	  \omega_1 \;\stackrel{\beta}{\longrightarrow}\; \omega_2
        \]
        and let $\pi=\beta\circ\alpha$
	be the composition of these maps. Then we have
	\begin{equation}\label{gl-12}
	   \Omega_\pi \;=\; \bigoplus_{k\in N}\;
	   \qquotient{\beta_q(\Omega_\alpha^k)}{\beta_q(J_{\beta_q}^k)}
        \end{equation}
	where $\Omega_\pi,\Omega_\alpha$ are given as in lemma
        \ref{lemm-2}, $\beta_q$
         is given by definition
	\ref{defi-1} and
        \[
            J_{\beta_q}^k = d_{\Omega_\alpha}\xkern \beta_q^{k-1} + \xkern
           \beta_q^{k}\;\;.
        \]
\end{prop}
{\bf Proof:}
	We shall go through the following proof in two steps. First the vector
        space isomorphism is established at the level of the $k$th grading.
        Then we show that
        multiplication and differential are respected thus giving an
        isomorphism of differential algebras.\newline
	Now $\xkern \alpha \subset \xkern \pi$ implies $J_\alpha\subset J_\pi$,
	hence we get
	\begin{equation}\label{gl-11}
	    \Omega_\pi^k
	    \;=\;
	      \qquotient{\pi(\omega^k)}{\pi(J_\pi^k)}=
	      \qquotient{\pi(\omega^k)}{\pi(J_\alpha^k)}
	      \bigg/
	      \qquotient{\pi(J_\pi^k)}{\pi(J_\alpha^k)}
	    \;=\;
      \beta_q\left( \qquotient{\alpha(\omega^k)}{\alpha(J_\alpha^k)}\right)
	      \bigg/
     \beta_q\left(\qquotient{\alpha(J_\pi^k)}{\alpha(J_\alpha^k)}\right)
	\end{equation}
	by definition of $\beta_q$. Next we show that
	\begin{equation}\label{gl-9}
	    \qquotient{\alpha(J_\pi^k)}{\alpha(J_\alpha^k)}=d_{\Omega_\alpha}
            \xkern \beta_q^{k-1}+\xkern \beta_q^k
	\end{equation}
	and therefore we first prove the identity
	\begin{equation}\label{gl-5}
	    \xkern\beta_q^k=
         \qquotient{\alpha(\xkern \pi^k +J_\alpha^k)}{\alpha(J_\alpha^k)}\;\;.
	\end{equation}
	Clearly ''$\supseteq$'' in equation \ref{gl-5} is given. On the
        other hand an element of
	$\xkern \beta_q^k$ is represented by $\nu\in\alpha(\omega^k)$ with
        $\beta(\nu)\in
	\pi(J_\alpha^k)$. This means $\beta(\nu)=\beta\circ\alpha(j)$ for an
        element
	$j\in J_\alpha^k$ and $\nu=\alpha(\rho)$ for some $\rho\in\omega$. Now
	$\nu=\alpha(\rho-j)+\alpha(j)$ with $\pi(\rho-j)=0$. Therefore
        $\nu\in\alpha(\xkern\pi^k+J_\alpha^k)$. This gives equation
        (\ref{gl-5}). Equation (\ref{gl-9}) is derived by :
	\begin{eqnarray*}
	   d_{\Omega_\alpha}\xkern \beta_q^{k-1}+\xkern \beta_q^k
	   &=&
	   d_{\Omega_\alpha}\left(
	     \qquotient{\alpha(\xkern\pi^{k-1}+
             J_\alpha^{k-1})}{\alpha(J_\alpha^{k-1})}
	   \right)+
             \qquotient{\alpha(\xkern\pi^k+J_\alpha^k)}{\alpha(J_\alpha^k)}\\
	   &=&\qquotient{\alpha(d \xkern\pi^{k-1})}{\alpha(J_\alpha^k)}+
   \qquotient{\alpha(\xkern\pi^k+d \xkern^{k-1}\alpha)}{\alpha(J_\alpha^k)}\\
 &=&\qquotient{\alpha(d \xkern\pi^{k-1}+\xkern\pi^k)}{\alpha(J_\alpha^k)}
	      \;=\;
	      \qquotient{\alpha(J_\pi^k)}{\alpha(J_\alpha^k)} \;\; .
	\end{eqnarray*}
	On the other hand
        \[
          \Omega_\alpha^k=\qquotient{\alpha(\omega^k)}{\alpha(J_\alpha^k)}\;\;
,
        \]
	such that equation \ref{gl-11} now reads
        \[
         \Omega_\pi^k =\qquotient{\beta_q(\Omega_\alpha^k)}{\beta_q(
	J_{\beta_q}^k)}\quad .
        \]
	This proves the vector space identity. Going through the proof so
        far it becomes clear that the isomorphism $i$ given by equation
        (\ref{gl-12}) is the identity map on representatives in
        $\pi(\omega)$ followed by different quotient building mechanisms.
        The quotient in $\Omega_\pi$ is split up into a double quotient.
	Using the definition of $d_{\Omega_\pi}$ on $\Omega_\pi$ and
        $d_{\sss RHS}$ given on
	\[
         \bigoplus_{k\in N}\qquotient{\beta_q(\Omega_\alpha^k)}{
          \beta_q(J_{\beta_q}^k)}
        \]
	by
        \[
           d_{\sss RHS}\cl{\cl{\nu}_{J_{\alpha}} }_{\beta_q(J_{\beta_q})} =
           \cl{d_{\Omega_\alpha}\cl{\nu}_{J_\alpha}}_{\beta_q(J_{\beta_q})}=
           \cl{\cl{\alpha(d\nu)}_{J_\alpha}}_{\beta_q(J_{\beta_q})}
        \]
        we have
        \[
           i\circ d_{\Omega_\pi}=d_{\sss RHS}\circ i \;\;.
        \]
        The corresponding relation holds for the multiplication defined on
        representatives in a similar fashion:
        \[
           \cl{\nu_1}_{\Omega_\pi}\circ \cl{\nu_2}_{\Omega_\pi}=
           \cl{\nu_1\nu_2}_{\Omega_\pi}
        \]
        \[
           \cl{\cl{\nu_1}_{J_\alpha}}_{\beta_q(J_{\beta_q})}\cdot
           \cl{\cl{\nu_2}_{J_\alpha}}_{\beta_q(J_{\beta_q})}
             =\cl{\cl{\nu_1\nu_2}_{J_\alpha}}_{\beta_q(J_{\beta_q})}
        \]
        such that
        \[
           i(\cl{\nu_1}\cl{\nu_2})=i(\cl{\nu_1})\cdot i(\cl{\nu_2})\quad .
        \]
        By these rules for multiplication and differential we thus have
        established an isomorphism of differential algebras.

We now apply proposition \ref{prop-1} to our situation.

\begin{theo}\label{theo-1}
   \begin{equation}\label{gl-2}
      \OD(\cA_1\otimes \cA_2)= \bigoplus_{k\in N}\qquotient{
      \pi_{\Sigma_q}\left( (\OD\cA_1\hat{\otimes}\OD\cA_2)^k\right)}
{\pi_{\Sigma_q}\left(d\xkern \pi_{\Sigma_q}^{k-1}+\xkern \pi_{\Sigma_q}^k
\right)}
   \end{equation}
\end{theo}
{\bf Proof:}
In proposition \ref{prop-1} we set
\[
 \omega=\Omega{\cA_1}\hat{\otimes}\Omega{\cA_2}
\]
which is justified by lemma \ref{lemm-2}. For
the maps we insert
\[
\alpha=\pi_\oplus\hspace{1cm}\beta=\pi_\Sigma\;\; .
\]
Using the result \ref{gl-15} it follows immediately that $\Omega_\pi$
is isomorphic to $\OD(\cA_1\otimes \cA_2)$.

For all examples of interest for particle physics the algebra has the
following form
\[
   {\cA}={\cF}\otimes{\cAM}\quad ,
\]
that is $\cA_1=\cF$ is the algebra of smooth functions on a manifold and
$\cA_2=\cAM$ is a matrix  algebra. For this special case the next
lemma shows, that the denominator in (\ref{gl-2}) vanishes. It
is not necessary to go as far as theorem \ref{theo-1} since it suffices
to use:
\[
    \OD(\cA_1\otimes\cA_2)^k
    \;=\;
    \qquotient{\pi(\Omega(\cA_1\hat{\otimes}\Omega\cA_2)^k)}{\pi(J_\oplus^k)}
    \bigg/
    \qquotient{\pi(J_\pi^k)}{\pi(J_\oplus^k)}
\]
If $\pi_1:{\cF} \rightarrow \cL(\cH_1)$ is a representation of
smooth functions on the square-integrable spinors as described in
\cite{cobuch},
$\pi_2:{\cA_2}\rightarrow M_n(C)$ an injective representation of a
matrix algebra $\cA_2$ on $\C^n$ we can construct a representation of
$\Omega\cF\hat{\otimes}\Omega\cAM$ on $\cH_1\otimes \C^n$ using the derivation
\[
   \cal D=[D,\cdot]+[\gamma^5\otimes \cM,\cdot]
\]
with the Dirac operator
$D=i\partial_\mu\gamma^\mu$ and an $n\times n$-matrix $\cM$. $\cA_2$ and
$\cM$ should be chosen in such a way that the condition (\ref{gl-22}) is
satisfied.
\begin{lemm}\label{lemm-1}
With the above preliminaries
\[
   \qquotient{\pi(J_\pi^k)}{\pi(J_\oplus^k)}=\{0\}
\]
\end{lemm}
{\bf Proof:}
During this proof we shall use the following shorthands:
\begin{equation}\label{gl-7}
   \pi_1(\Omega\cF^k)=\omega_1^k\qquad\pi_2(\Omega\cAM^k)
   =\omega_2^k\qquad\pi_i(J_i^k)=j_i^k
\end{equation}
The important property of $\Omega(\cF)$ is for $k\geq2$:
\[
    j_1^k=\omega_1^{k-2} \subseteq \omega_1^{k}
\]
Thus we have
\begin{eqnarray}
   \pi(J_\oplus^k)&=& \sum_{i+j=k} j_1^i\otimes\omega_2^j +
   \omega_1^i\otimes j_2^j\nonumber\\
  &=&
sum_{i+j=k-2}\omega_1^i\otimes\omega_2^j+\sum_{i+j=k}\omega_1^i\otimes j_2^j
\;\;.
   \label{gl-10}
\end{eqnarray}
We shall now choose a representative $\alpha\in \pi(J_\pi^k)$ and show that
$\alpha\in\pi(J_\oplus^k)$.
$\alpha$ can be written as $\alpha=\pi(\delta k)$ with
\begin{equation}\label{gl-8}
   k=\bigoplus_{i+j=k-1} k_1^i\otimes k_2^j\quad\in\xkern\pi^{k-1}
\end{equation}
and
\[
   k_1^i = f_0^i\delta_1f_1^i\ldots \delta_1f_i^i
   \quad,\quad
   k_2^j= A_0^j\delta_2A_1^j\ldots \delta_2A_j^j \;\;.
\]
where $f$ are functions and $A\in\cAM$. In equation (\ref{gl-8}) we
have suppressed a further summation due to the tensor product in order to
simplify notation. Then
\begin{eqnarray}
      \alpha &=& \pi(\delta k) \;=\;
      \sum_{i+j=k-1}
      \pi_1(\delta_1k_1^i)(\gamma^5)^j\otimes\pi_2(k_2^j)+(-1)^i
      \pi_1(k_1^i)(\gamma^5)^{j+1}\otimes\pi_2(\delta_2k_2^j) \nonumber \\[1ex]
      &=&
      \sum_{i+j=k-1}
    \dD f_0^i\dD f_1^i\ldots \dD f_i^i(\gamma^5)^j\otimes A_0^j\dM A_1^j\ldots
      \dM A_j^j\nonumber\\[1ex]
      & &
      + \sum_{i+j=k-1} (-1)^i f_0^i\dD f_1^i\ldots
      \dD f_i^i(\gamma^5)^{j+1}\otimes \dM  A_0^j\dM A_1^j\ldots
      \dM A_j^j
   \label{gl-13}
\end{eqnarray}
with
\[
   \dD f=[D,f]=\pi_1(\delta_1f)
   \quad,\quad
   \dM A=[\cM,A]=\pi_2(\delta_2A) \;\;.
\]
Since $k\in\xkern\pi^{k-1}$ we also have
\begin{equation}\label{gl-3}
   0=\sum_{i+j=k-1}f_0^i\dD f_1^i\ldots \dD f_i^i(\gamma^5)^j\otimes
   A_0\dM A_1^j\ldots \dM A_j^j \;\;.
\end{equation}
We shall first deal with the second term of the r.h.s. of (\ref{gl-13}).
We know from section 2 that $\dM $ can be written as a supercommutator up to
elements generated by $[\cM^2,\cdot]$ which are in the ideal (\ref{gl-10}).
Thus we can rewrite this term as
\[
   \left[\gamma^5\otimes M, \sum_{i+j=k-1}f_0^i\dD f_1^i
   \ldots \dD f_i^i(\gamma^5)^j\otimes A_0\dM A_1^j\ldots \dM A_j^j \right]_S
   +\mbox{Terms in} [\cM^2,\cdot]
\]
and therefore it is contained in $\pi(J_\oplus^k)$.
We now turn to the first term of the r.h.s. of (\ref{gl-13}).
By using $\dD f=[D,f]$ and equation (\ref{gl-3})
we obtain
\[
   \mbox{first term of (\ref{gl-13})}
   \;=\;
   -\sum_{i+j=k-1}f_0^iD(Df_1^i)\ldots (Df_i^i)(\gamma^5)^j\otimes
   A_0\dM A_1^j\ldots \dM A_j^j \;\;.
\]
At least two of the $\gamma$-matrices appearing in this expression
are identical, therefore this term is contained in
\[\sum_{i+j=k-2}\omega_1^i\otimes\omega_2^j\]
and therefore in $\pi(J_\oplus^k)$ according to (\ref{gl-10}). This gives
$\alpha\in\pi(J_\oplus^k)$, thus completing the proof.

The results of this section can be summarized as follows:

For all algebras $\cA = \cF\otimes \cAM$ fulfilling the preliminaries of
lemma
\ref{lemm-1}, that is for all algebras relevant for particle physics,
$\OD$ is given by (using the conventions \ref{gl-7})
\begin{equation}
   \OD(\cF\otimes \cAM) =
   \frac{\omega_1^k\otimes\omega_2^0+\ldots+\omega_1^0\otimes\omega_2^k}
     {
      \omega_1^{k-2}\otimes\omega_2^0+\ldots+\omega_1^0\otimes\omega_2^{k-2}+
      \omega_1^{k-2}\otimes j_2^2+\ldots+\omega_1^0\otimes j_2^k
     }\label{gl-6}
\end{equation}
This can easily be calculated for any specific example once one has
determined the relevant terms for $\cF$ and $\cAM$.


\section{The Two Point Case}
In this section we want to apply the general results, developed in the
previous section, to the two point case. The algebra $\cA$ is
\[
   \cA = \cF\otimes (\cA_1\oplus\cA_2) =\cF\otimes\cAM
\]
where $\cA_1$, $\cA_2$ denote $\C^{m\times m}$ resp.~$\C^{n\times n}$
matrix algebras. The Dirac operator for $\Omega(\cA_1\oplus\cA_2)$ is
off-diagonal as in (\ref{gl-22}). Thus $\OD(\cA_1\oplus\cA_2)$ is
known and we can distinguish  three different cases as shown at the end
of sec.~3.
$\OD\cF$ is the de-Rham complex \cite{cobuch} and the Dirac operator
for the product k-cycle is
\[
   {\cal D} = i\dslash\otimes 1 + [\gamma^5\otimes\cM , \cdot ]\; ,
\]
A general result is
\[
\OD\cA^0 = \cA \;\; ,\;\; \OD\cA^1=\big(\OD\cF\hat\otimes\OD(\cA_1\oplus\cA_2)
\big)^1
\]
as one immedeately infers from equation (\ref{gl-6}).

We now analyze the higher degrees of $\OD\cA$ for the three different
$\OD(\cA_1\oplus\cA_2)$ as in sec.~3.
\begin{itemize}
\item[{\bf i.}] $\mu^*\mu \sim 1_{m\times m}$ and $\mu\mu^*
   \sim 1_{n\times n}$, i.e.
   $\cA_1=\cA_2$. Because of the isomorphism (\ref{gl-16}) we have:
   \[
      \pi_1(\Omega\cF^k)\otimes\pi_2(\Omega\cAM^{l+2}) =
      \pi_1(\Omega\cF^k)\otimes\pi_2(\Omega\cAM^{l})
   \]
   Inserting this in equation (\ref{gl-6}) and using the fact that
   \[
      \pi_1({\cJ_{\cF}}^k)=\pi_1(\Omega\cF^{k-2})
   \]
   and
   \[
      \pi_2(\cJ_{\cAM}^k)=\{0\}
   \]
   one obtains for $k=2$
   \begin{eqnarray*}
         \OD\cA^2
         &=&
         \frac{\pi_1(\Omega\cF^2)\otimes\pi_2(\Omega\cAM^0) +
               \pi_1(\Omega\cF^{1})\otimes\pi_2(\Omega\cAM^1)}
              {\pi_1(\Omega\cF^{0})\otimes\pi_2(\Omega\cAM^0)} \\
         &=&
         \Lambda^2\otimes\cAM +\Lambda^1\otimes\cM\cAM
   \end{eqnarray*}
   and for $k>2$:
   \begin{eqnarray*}
         \OD\cA^k
         &=&
         \frac{\pi_1(\Omega\cF^k)\otimes\pi_2(\Omega\cAM^0) +
               \pi_1(\Omega\cF^{k-1})\otimes\pi_2(\Omega\cAM^1)}
              {\pi_1(\Omega\cF^{k-2})\otimes\pi_2(\Omega\cAM^0) +
               \pi_1(\Omega\cF^{k-3})\otimes\pi_2(\Omega\cAM^1)}\\
         &=&
         \Lambda^k\otimes\cAM +\Lambda^{k-1}\otimes\cM\cAM
   \end{eqnarray*}
In the quotient we suppressed further terms which trivially cancel out.
$\Lambda^k$ denotes the space of differential forms of degree $k$.
The degree of an element $\alpha \in \OD\cA$ is the sum of the form degree
and the degree of the matrix algebra. We see that although all degrees
$k\in\N$ of the matrix algebra $\OD\cAM$ are non-trivial, in the algebra
$\OD\cA$ only the zeroth and first matrix degrees appear. However, this
situation can change if we include a "generation-space", i.e. if we
allow for a bigger representation space for the algebra $\cAM$. The
homomorphism
\[
   \pi_2 :\cAM\lra\C^{2m \times 2m}
\]
is extended to
\[
   \pi^\prime_2 : \cAM\lra\C^{2m \times 2m}\otimes\C^{g\times g}
\]
where $\C^g$ is the "generation-space". The new homomorphism is given as
\[
   \pi^\prime_2 = \pi_2\otimes 1
\]
This by itself  would not yield higher matrix degrees in $\OD\cAM$. In order
to get that we
have to use the larger freedom in the choice of the matrix $\cM$. We now take
\[
   \cM^\prime = \left(\begin{array}{cc}
                      0         & \mu^*\otimes G^* \\
                   \mu\otimes G &  0
                \end{array}\right)\;\; .
\]
Here $G$ denotes an arbitrary $\C^{g\otimes g}$-matrix. The effect of this
extension is that we now can distinguish between elements
$\alpha \in \pi_1(\Omega\cF^k\otimes\Omega\cAM^p)$ and
$\beta \in \pi_1(\Omega\cF^k\otimes\Omega\cAM^q)$  for $p\neq q$ as
long as the powers of $G^*G$ resp.~$GG^*$ are lineary independent for $p$
and $q$. Therefore they cannot be cancelled by the denominator of
equation (\ref{gl-6}). However there is an integer $p_0\leq g$ for
which the powers of the matrices become linearly dependent. In this case
any element $\alpha_{p_0} \in
 \pi(\Omega\cF^k\otimes\Omega\cAM^{p_0})$ can be written as
a linear combination of elements with smaller matrix degree:
\[
   \alpha_{p_0} = \sum_{q=1}^{q\leq p_0/2} \alpha_{p_0-2q}\;\;\; ,\;\;
   \alpha_{p_0-2q} \in \pi(\Omega\cF^k\otimes\Omega\cAM^{p_0-2q})
\]
As a consequence all terms with matrix degree $p\geq p_0$ in $\pi(\Omega\cA)$
are cancelled by the denominator of (\ref{gl-6}).

These results can be summarized by defining the following representation
for $\OD\cA$. The matrix part $\cAM^\prime$ of the algebra is generated by the
zeroth order
\[
   \cA^\prime_0 = \left(\begin{array}{cc}
   \cA_1 & 0 \\
   0   & \cA_1
   \end{array}\right)
\]
and the first order
\[
   \cA^\prime_1 = \left(\begin{array}{cc}
   0 & \eta\cA_1 \\
   \eta\cA_1   & 0
   \end{array}\right)\;\; .
\]
Here we have introduced a formal element $\eta$ which has the property that
\[
   \eta^{p_0} = 0\;\;\; ;\;\;\; \eta^p \neq 0\; , p<p_0
\]
and it commutes with all other elements of the algebra. Thus
$\cAM^\prime$
is a graded algebra with highest degree $p_{0}-1$ and an induced $\Z_2$
grading. The full algebra $\OD\cA$ is obtained by taking the graded tensor
product of
$\cAM^\prime$ and the de Rham algebra $\Lambda$
\[
   \OD\cA = \Lambda\hat{\otimes}\cAM^\prime\;\; .
\]
The degree of elements in $\OD\cA$ is the sum of form degree and matrix degree.
The derivation on an element $\alpha \in \OD\cA^k$ is given as
\[
   d\alpha = d_C\alpha + \left[
   \left(\begin{array}{cc}
      0    & \eta \\
      \eta & 0 \end{array}\right) , \alpha \right]
\]
where $d_C$ denotes the usual exterior derivative and the commutator
is the graded commutator.

One now might wonder what has happened to the "generation" space? In fact,
it is not needed at the level of the algebra since it was introduced to
separate the matrix degrees. This task has been taken over by the element
$\eta$ which also has the nilpotency property $\eta^{p_0}=0$. However,
the "generation"-matrix $M$ may have a physical interpretation as a mass-matrix
for fermions and one might wish to keep it in the algebra. This is of
course possible and does not change any algebraic properties.

\item[{\bf ii.}] $\mu^*\mu \sim\!\!\!\!\!\!/\; 1_{m\times m}$ and
$\mu\mu^* \sim\!\!\!\!\!\!/\; 1_{n\times n}$. In this case $\OD^1\cAM$ is
the highest matrix degree in $\OD\cF\hat{\otimes}\OD\cAM$ and therefore
no further cancellations appear in equation (\ref{gl-6}), i.e.,
\[
\mbox{ker}\pi_{\Sigma_q} = \{0\}\;\; .
\]
Thus we infer that
\[
   \OD\cA = \OD\cF\hat{\otimes}\OD\cAM =\Lambda\hat{\otimes}\OD\cAM
\]
Note that the introduction of a
"generation"-space would not change the situation.

\item[{\bf iii.}] $m \leq n$, $\mu^*\mu \sim 1_{m\times m}$ and
$\mu\mu^* \sim\!\!\!\!\!\!/\; 1_{n\times n}$. In this case the highest matrix
degree is 2 but
\[
   \pi(\Omega\cF^k\otimes\Omega\cAM^2) \subset
   \pi_2(\Omega\cF^k\otimes\Omega\cAM^0)\;\; .
\]
Therefore the highest matrix degree in $\OD\cA$ is 1 and we can represent
this algebra in the same way as in {\bf ii.}. The extension by a
"generation"-space as in {\bf i.} can  be used to make
$\pi(\Omega\cF^k\otimes\Omega\cAM^2)$ and
$\pi_2(\Omega\cF^k\otimes\Omega\cAM^0)$
distinguishable. In this case we again have
\[
   \mbox{ker}\pi_{\Sigma_q} = \{0\}\;\;
\]
and therefore
\[
   \OD\cA =\OD\cF\hat{\otimes}\OD\cAM =\Lambda\hat{\otimes}\OD\cAM\;\; .
\]
\end{itemize}


\section{Conclusions}
We derived a general formula which relates the differential algebra
$\OD(\cA_1\otimes\cA_2)$ of a product algebra to the differential algebras
$\OD\cA_1$ and $\OD\cA_2$ of the factor algebras. This considerably simplifies
the calculation of $\OD(\cA_1\otimes\cA_2)$ once the differential algebras of
the factor algebras are known.

However, in the context of Yang-Mills theories with spontaneous symmetry
breaking, all relevant algebras are of the form $\cA=\cF\otimes\cAM$ with
$\cF$, the algebra of smooth functions on space-time and $\cAM$, a
matrix-algebra. In this case the differential algebras of each factor
algebra is known. For the algebra of functions it is the usual de
Rham-algebra \cite{cobuch} and the
differential algebras for matrix-algebras are described in sec.~3. With this
information it is possible to compute the full differential algebra
$\OD(\cF\otimes\cAM)$ as we showed for the two-point case in some detail.
The generalization to three or more points, necessary to
handle models with several symmetry breaking scales, is obvious although
it would require a new calculation  along the lines described in sec.~4.

Since physical models, at least the bosonic part, are constructed out of
objects in $\OD\cA$, which is an $\cA$-module, the explicit knowledge of
the differential algebra for a given algebra $\cA$ allows for a very
economical derivation of physical quantities like connection and curvature.
This can be done in the usual way by taking an antihermitian one form as
connection form and the curvature as the square
of the connection. However, the construction of physically relevant models
requires a more careful discussion, e.g. the imbedding of charge and iso-spin
enforces a certain structure on the Higgs-sector. We shall come
back to this point in a future publication.

It is now also possible to use the explicit knowledge of $\OD\cA$ to discuss
the precise relation of Connes' approach to Yang-Mills theory with spontaneous
symmetry breaking and the model presented in \cite{robert,florian,hps2}. This
latter model is based on superconnection a la Matthai, Quillen \cite{MaQui}.
Here, the usual exterior differential is extended by a matrix differential,
connections
are elements of odd degree in a graded $SU(n|m)$ algebra extended to a module
over differential forms. There are several features in this approach similar
to the differential algebras derived in sec. 5, namely the general settings in
matrix valued differential forms, the matrix derivation and Cartans derivation
giving the building principles for connection and curvature. However, we also
note an important difference: the quotient building described in sec. 5 does
not occur there and therefore the model is not based on a differential
algebra, but an algebra with derivation.

So far, we have only discussed the bosonic sector of physical models. For the
derivation of $\OD\cA$ one has to introduce a Dirac operator in order to
represent the differential envelope. It is considered as one nice property
of Connes approach, that this Dirac operator can be used to write down the
fermionic Lagrangian. However, if one starts with $\OD\cA$ for the construction
of physical models, then there is no Dirac operator given automatically.
Of course, such a Dirac operator can be derived by requiring the usual
physical properties.




\begin{thebibliography}{99}
\bibitem{cobuch} A.~Connes,
        `Non Commutative Geometry',
	Academic Press, in press, IHES/M/93/12 (1993).
\bibitem{CoLo}A.~Connes, J.~Lott, Nucl.~Phys.~B Proc.~Supp.~18B (1990) 12.
\bibitem{chams} A.H.~Chamseddine, G.~Felder, J.~Fr\"ohlich,
        Nucl.~Phys.~B395 (1992) 672; Phys.~Lett.~B296 (1992) 109;\\
        A.H.~Chamseddine, J.~Fr\"ohlich,
        ZU-TH-10-1993, (1993).
\bibitem{chams1} A.H.~Chamseddine, J.~Fr\"ohlich,
        Phys.~Lett.~B314 (1993) 308.
\bibitem{robert} R.~Coquereaux, G.~Esposito-Far\`ese, G.~Vaillant,
        Nucl.~Phys. B353 (1991) 689;\\
        R.~Coquereaux, in `Differential Geometric Methods in Theoretical
        Physics'
        Lecture Notes in Physics 375, eds.~C.~Bartocci et al., Springer Verlag
        (1990) 3.
\bibitem{florian}R.~H\"au\ss{}ling, N.A.~Papadopoulos, F.~Scheck,
Phys.~Lett.~B260 (1991) 125;\\
        R.~Coquereaux, G.~Esposito-Far\`ese, F.~Scheck,
        Int.~J of Mod.~Phys., A7 (1992) 6555.
\bibitem{hps2} R.~H\"au{\ss}ling, N.A.~Papadopoulos, F.~Scheck,
        Phys. Lett. {\bf B303} (1993) 265.
\bibitem{kastler} D.~Kastler,
        'A detailed account of A. Connes` version of
        the Standard Model in Non-Comutative Geometry',
        CPT-91/P.2610, Marseille (1992).
\bibitem{KT} D.~Kastler, D.~Testard,
        'Quantum Forms of tensor products',
	CPT-92/P.2793.
\bibitem{pps} N.A.~Papadopoulos, J.~Plass, F.~Scheck,
        'Models of electroweak interactions in Non-Commutative Geometry:
        a comparison',
	MZ-TH/93-26.
\bibitem{josch} J.~Plass,
        'Verallgemeinerte Differentialformen als Zugang zur Nichtkommutativen
	Differentialgeometrie',
	Diplomarbeit, Mainz (1993).
\bibitem{MaQui} D.~Quillen,
        Topology 24, (1985) 89; \\
        V.~Matthai, D.~Quillen,
	Topology 25, (1986) 85.
\bibitem{jan} J.-M.~Warzecha,
        'Nichtkommutative Geometrie und das Standardmodell der
        Elementarteilchenphysik',
	Diplomarbeit, Mainz (1993).
\end{thebibliography}
\end{document}